\documentclass[a4paper,twocolumn]{article}
\usepackage[dvips]{graphicx}
\begin{document}

\title{Realization of quantum operations on photonic qubits
by linear optics and post-selection}

\author{Holger F. Hofmann
and Shigeki Takeuchi \\ PRESTO, Japan Science and
Technology Corporation (JST)\\  Research Institute for Electronic Science,\\ Hokkaido University, Sapporo 060-0812\\
Tel/Fax: 011-706-2648\\ 
e-mail: h.hofmann@osa.org}

\date{}

\maketitle

\begin{abstract}
One of the greatest difficulties in the applications of
single photon polarization states as qubits is 
the realization of controlled interactions between two
photons. Recently, it has been shown that such interactions
can be realized using only beam splitters and high efficiency
photon detection by post-selecting a well defined part of the
results in the output. We analyze these interactions and
discuss schemes for qubit operations based on this mechanism. 
\\[0.2cm]
\end{abstract}

\section{Introduction}
Conventionally, the unitary transformations representing
controlled quantum gates are implemented by appropriate
interactions between physical systems. These interactions
should be non-dissipative and involve only the four level
Hilbert space of the two interacting qubits. 
If photonic qubits are realized using the polarization
states of individual photons, such an interaction
requires a Kerr nonlinearity strong enough to cause a 
phase change of $\pi$ per photon in the controlling 
mode, while all interactions with modes other than the 
two well-defined polarization modes of each photon must be 
suppressed. However, the suppression of all absorption 
and scattering processes in the presence of a strong
Kerr nonlinearity is very difficult to achieve and requires
the development of highly sophisticated new technologies.
The implementation of controlled interactions between photonic
qubits has therefore been a major obstacle to the realization
of quantum computation using photonic qubits.

To overcome this obstacle, Knill, Laflamme and Milburn have
proposed an alternative scheme for quantum computation using
photonic qubits \cite{Kni01}. This alternative scheme is based
on the insight that the necessary unitary transforms needed
for controlled gate operations can also be obtained from a
subspace of a total unitary transformation in a much larger
Hilbert space. Although the selection of such a subspace 
corresponds to a limited probability of success for the 
gate operation, this problem can be compensated by applying 
various error correction strategies. 
In optics, this alternative realization of controlled gate
operations is especially promising because it allows an
implementation of photon-photon interactions using only 
linear optical elements. In particular, the selection of
appropriate subspaces in both the input and the output 
of a conventional beam splitter is sufficient to obtain
basic nonlinear interactions between photons.

\section{Properties of the post-selected beam splitter}
\begin{figure}
\begin{picture}(240,200)
%\put(0,0){\framebox(240,160){}}
\put(50,30){\line(1,1){20}}
\put(50,30){\line(-1,1){20}}
\put(30,50){\line(1,1){20}}
\put(50,70){\line(1,0){20}}
\put(70,50){\line(0,1){20}}
\put(43,41){\makebox(20,20){\large $\mid \! n_2 \rangle$}}
\put(20,10){\makebox(60,20){\large Photon Source}}
\put(70,70){\line(1,1){60}}
\put(130,130){\line(-2,-1){10}}
\put(130,130){\line(-1,-2){5}}
\put(110,135){\makebox(100,20){
\large $\mid \! \psi_{\mbox{out}} \rangle = 
\hat{S}_{nn}\mid \!\psi_{\mbox{in}} \rangle$}}

\put(70,100){\line(1,0){60}}
\put(130,100){\makebox(50,15){\large Beam}}
\put(130,85){\makebox(50,15){\large Splitter}}

\put(70,130){\line(1,-1){65}}
\put(135,65){\line(-2,1){10}}
\put(135,65){\line(-1,2){5}}
\put(40,135){\makebox(40,20){
\large $\mid \! \psi_{\mbox{in}} \rangle$}}

\put(130,50){\line(1,1){20}}
\bezier{100}(130,50)(140,40)(150,50)
\bezier{100}(150,50)(160,60)(150,70)
\bezier{100}(150,50)(160,40)(170,50)

\put(170,50){\makebox(20,20){\large $\langle n_2 \! \mid$}}
\put(130,18){\makebox(60,20){\large Photon Counter}}

\end{picture}
\caption{\label{basic} Schematic representation of the
elements for a realization of basic post-selected beam
splitter operations.}
\end{figure}
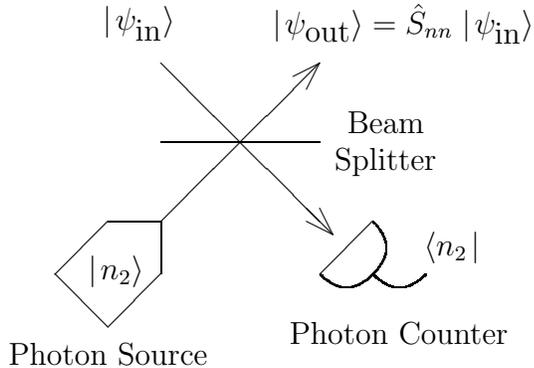

The action of a beam splitter of reflectivity $R$ on the 
two input modes given by the operators $\hat{a}_1$ and 
$\hat{a}_2$ can be described by a unitary transformation 
$\hat{U}_R$ with the property
\begin{eqnarray} 
\hat{U}_R \; \hat{a}_1 \hat{U}_R^\dagger 
&=& \sqrt{R} \; \hat{a}_1 + i \sqrt{1-R} \; \hat{a}_2
\nonumber \\
\hat{U}_R \; \hat{a}_1 \hat{U}_R^\dagger 
&=& i \sqrt{1-R} \; \hat{a}_1 + \sqrt{R} \; \hat{a}_2.
\end{eqnarray}
While the total number of photons is a conserved quantity
of this operation, photons are exchanged between mode 1 
and mode 2, so that most of the output results change the
photon numbers in the modes. However, the concept of photonic
qubits is based on the assumption that the photons can be
kept in separate modes. Therefore, the distribution of photons
at the beam splitter must be controlled by post-selecting only
those output results that conserve the photon number 
distribution. 
Since the beam splitter automatically conserves the total 
photon number, it is sufficient to ensure that the photon
number on only one side of the beam splitter does not change.
This strategy can be implemented by injecting a photon number
state $\mid \! n_2\rangle$ in mode 2 and detecting the same 
output photon number $\langle n_2 \! \mid$ in the output of
mode 2. Figure \ref{basic} illustrates this basic concept.
For the most simple case of a vacuum input and no photon
in the detector output ($n_2=0$), this device causes a linear
attenuation of the input amplitudes, as given by the 
effective operator elements
\begin{equation}
\label{eq:s00}
\langle n_1; 0 \! \mid \hat{U}_R \mid \! n_1; 0 \rangle 
= \langle n_1 \! \mid \hat{S}_{00} \mid \! n_1 \rangle 
= \left(\sqrt{R}\right)^{n_1}.
\end{equation}
This non-unitary beam splitter operation
can be used to compensate amplitude differences between
the photon number states of mode 1.
However, the most important result is obtained for a single 
photon input and a single photon detection ($n_2=1$).
This device produces a nonlinear effect given by the effective
operator elements
\begin{eqnarray}
\label{eq:s11}
\lefteqn{
\langle n_1; 1 \! \mid \hat{U}_R \mid \! n_1; 1 \rangle 
= \langle n_1 \! \mid \hat{S}_{11} \mid \! n_1 \rangle} 
\nonumber \\
&& \hspace{0.5cm}
= \left(\sqrt{R}\right)^{n_1-1}\left( R-(1-R)\; n_1 \right).
\end{eqnarray}
This matrix element is positive for all $n_1<R/(1-R)$ and
negative for all $n_1>R/(1-R)$. Therefore, equation 
(\ref{eq:s11}) describes a nonlinear phase change of $\pi$ 
at an intensity of $R/(1-R)$ photons. In particular, a 
nearly unitary result is obtained in the subspace of
$n_1\leq 2$ for $R=1/4$, where 
\begin{eqnarray}
\hat{S}_{11}(R=1/4) \mid\! 0 \rangle &=& 
\frac{1}{2}\mid\! 0 \rangle
\nonumber \\
\hat{S}_{11}(R=1/4) \mid\! 1 \rangle &=& 
-\frac{1}{2}\mid\! 1 \rangle
\nonumber \\
\hat{S}_{11}(R=1/4) \mid\! 2 \rangle &=& 
-\frac{5}{8}\mid\! 2 \rangle.
\end{eqnarray}
Both the original scheme proposed by Knill, Laflamme and Milburn
\cite{Kni01}
and the simplified proposal of Ralph, White, Munro and Milburn
\cite{Ral02} use reflectivities close to $R=1/4$ to realize
their basic nonlinearity. By choosing a reflectivity slightly
higher than $1/4$, the two proposals accomplish a gradual 
increase in the output amplitude from zero photons to two photons.
This increase can then be compensated by linear attenuation.
In the proposal of Ralph et al. \cite{Ral02}, this is done
by using the linear attenuation of the vacuum beam splitter 
given in equation (\ref{eq:s00}) to compensate an effective linear
amplification obtained from the single photon operation of
equation (\ref{eq:s11}).  

While the basic operation of the beam splitter thus allows the
realization of well defined nonlinear interactions between
photons, the overall effort required to adjust the amplitudes
of the output is still significant. Therefore, it may be 
worthwhile to consider applications where this kind of adjustment
is less critical.

\section{Nonlinear filter operation}
Instead of compensating the amplitude differences between the
photon number states caused by the different matrix elements
of $\hat{S}_{11}$, the nonlinear features of these amplitude 
differences may be exploited directly in order to filter out 
specific components of the multi photon quantum state
\cite{Hof02}.
The most fundamental quantum filter is then obtained by the
beam splitter with reflectivity $R=1/2$ - the standard type
of beam splitter used in most quantum optics experiments.
Its post-selection properties are given by
\begin{eqnarray}
\hat{S}_{11}(R=1/2) \mid\! 0 \rangle &=& 
\frac{1}{\sqrt{2}}\mid\! 0 \rangle
\nonumber \\
\hat{S}_{11}(R=1/2) \mid\! 1 \rangle &=& 0
\nonumber \\
\hat{S}_{11}(R=1/2) \mid\! 2 \rangle &=& 
-\frac{1}{2 \sqrt{2}}\mid\! 2 \rangle.
\end{eqnarray}
This operation eliminates only the one photon component
of the input. Both the vacuum and the two photon component 
can pass this filter without any loss of quantum coherence.

In order to apply this filtering process to photonic
qubits, it is necessary to temporarily transfer the
photons of one polarization component of each qubit to
the same optical mode. This is usually accomplished by 
reversible photon bunching in a Mach-Zender geometry.
Specifically, we can apply the reversible transformation of
an additional pair of beam splitters with reflectivity $1/2$
to realize the following operation on a two mode input:
\small
\begin{eqnarray}
\label{eq:core}
\hat{U}_{1/2}(1,2)
\;(\hat{S}_{11}(1)\otimes\hat{S}_{11}(2))\; \hat{U}_{1/2}(1,2) 
\mid\! 1;1 \rangle &=&
\hspace{-0.3cm}
 -\frac{1}{4} \mid\! 1;1 \rangle 
\nonumber \\
\hat{U}_{1/2}(1,2)
\;(\hat{S}_{11}(1)\otimes\hat{S}_{11}(2))\; \hat{U}_{1/2}(1,2) 
\mid\! 0;1 \rangle &=& 0
\nonumber \\[0.1cm]
\hat{U}_{1/2}(1,2)
\;(\hat{S}_{11}(1)\otimes\hat{S}_{11}(2))\; \hat{U}_{1/2}(1,2) 
\mid\! 1;0 \rangle &=& 0 
\nonumber \\
\hat{U}_{1/2}(1,2)
\;(\hat{S}_{11}(1)\otimes\hat{S}_{11}(2))\; \hat{U}_{1/2}(1,2) 
\mid\! 0;0 \rangle &=& \hspace{-0.1cm}
\frac{1}{2} \mid\! 0;0 \rangle.
\nonumber \\
\end{eqnarray}
\normalsize
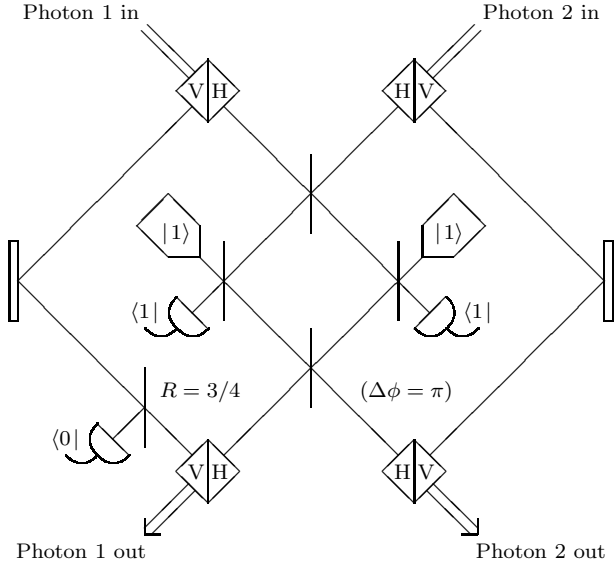
\begin{figure}
\footnotesize
\setlength{\unitlength}{0.6pt}
\begin{picture}(400,400)
%\put(0,0){\framebox(400,400){}}

%====V1

%--photon1in
\put(123,328){\line(-1,1){30}}
\put(127,332){\line(-1,1){30}}
\put(15,360){\makebox(80,20){Photon 1 in}}
%--polarizer1in
\put(135,300){\line(0,1){40}}
\put(135,340){\line(1,-1){20}}
\put(135,340){\line(-1,-1){20}}
\put(135,300){\line(1,1){20}}
\put(135,300){\line(-1,1){20}}
\put(120,310){\makebox(15,20){V}}
\put(135,310){\makebox(15,20){H}}
%--V1 path
\put(15,200){\line(1,-1){110}}
\put(10,175){\framebox(5,50){}}
\put(15,200){\line(1,1){110}}
%--polarizer1out
\put(135,60){\line(0,1){40}}
\put(135,100){\line(1,-1){20}}
\put(135,100){\line(-1,-1){20}}
\put(135,60){\line(1,1){20}}
\put(135,60){\line(-1,1){20}}
\put(120,70){\makebox(15,20){V}}
\put(135,70){\makebox(15,20){H}}
%--photon1out
\put(123,72){\line(-1,-1){28}}
\put(127,68){\line(-1,-1){28}}
\put(95,40){\line(1,0){10}}
\put(95,40){\line(0,1){10}}
\put(15,20){\makebox(80,20){Photon 1 out}}

%====V2

%--photon2in
\put(277,328){\line(1,1){30}}
\put(273,332){\line(1,1){30}}
\put(305,360){\makebox(80,20){Photon 2 in}}
%--polarizer2in
\put(265,300){\line(0,1){40}}
\put(265,340){\line(-1,-1){20}}
\put(265,340){\line(1,-1){20}}
\put(265,300){\line(-1,1){20}}
\put(265,300){\line(1,1){20}}
\put(250,310){\makebox(15,20){H}}
\put(265,310){\makebox(15,20){V}}
%--V2 path
\put(385,200){\line(-1,-1){110}}
\put(385,175){\framebox(5,50){}}
\put(385,200){\line(-1,1){110}}
%--polarizer2out
\put(265,60){\line(0,1){40}}
\put(265,100){\line(-1,-1){20}}
\put(265,100){\line(1,-1){20}}
\put(265,60){\line(-1,1){20}}
\put(265,60){\line(1,1){20}}
\put(250,70){\makebox(15,20){H}}
\put(265,70){\makebox(15,20){V}}
%--photon2out
\put(277,72){\line(1,-1){28}}
\put(273,68){\line(1,-1){28}}
\put(305,40){\line(-1,0){10}}
\put(305,40){\line(0,1){10}}
\put(305,20){\makebox(80,20){Photon 2 out}}

%====H12

%--upper and lower beam splitters
\put(200,230){\line(0,1){50}}
\put(200,120){\line(0,1){50}}

%--right input path
\put(145,310){\line(1,-1){130}}
%--right beam splitter
\put(255,175){\line(0,1){50}}
%--right detector
\put(265,170){\line(1,1){20}}
\bezier{80}(265,170)(275,160)(285,170)
\bezier{80}(285,170)(295,180)(285,190)
\bezier{80}(285,170)(295,160)(305,170)
\put(285,170){\makebox(40,20){$\langle 1\!\mid$}}

%--left input path
\put(255,310){\line(-1,-1){130}}
%--left beam splitter
\put(145,175){\line(0,1){50}}
%--left detector
\put(135,170){\line(-1,1){20}}
\bezier{80}(135,170)(125,160)(115,170)
\bezier{80}(115,170)(105,180)(115,190)
\bezier{80}(115,170)(105,160)(95,170)
\put(75,170){\makebox(40,20){$\langle 1 \!\mid$}}

%--right output path
\put(270,215){\line(-1,-1){125}}
%--right photon source
\put(270,215){\line(1,0){20}}
\put(270,215){\line(0,1){20}}
\put(290,255){\line(-1,-1){20}}
\put(310,235){\line(-1,-1){20}}
\put(310,235){\line(-1,1){20}}
\put(278,221){\makebox(18,18){$\mid\! 1 \rangle$}}

%--left output path
\put(130,215){\line(1,-1){125}}
%--left photon source
\put(130,215){\line(-1,0){20}}
\put(130,215){\line(0,1){20}}
\put(110,255){\line(1,-1){20}}
\put(90,235){\line(1,-1){20}}
\put(90,235){\line(1,1){20}}
\put(106,221){\makebox(18,18){$\mid\! 1 \rangle$}}

%==VV-attenuation
%--3/4 beam splitter 
\put(95,95){\line(0,1){50}}
\put(110,120){\makebox(40,20){$R=3/4$}}
%--subtracted line & detector
\put(95,120){\line(-1,-1){20}}
\put(85,90){\line(-1,1){20}}
\bezier{80}(85,90)(75,80)(65,90)
\bezier{80}(65,90)(55,100)(65,110)
\bezier{80}(65,90)(55,80)(45,90)
\put(25,90){\makebox(40,20){$\langle 0 \!\mid$}} 

%==phase correction

\put(230,120){\makebox(60,20){($\Delta \phi = \pi$)}}

\end{picture}
\normalsize
\setlength{\unitlength}{1pt}
\vspace{-0.5cm}
\caption{\label{filter} Schematic setup of the quantum filter
for two photon polarization correlations. Unless labeled
otherwise, vertical lines represent beam splitters with 
reflectivity $R=1/2$. The boxes labled with H and V represent
polarization sensitive beam splitters transmitting H polarized
photons and reflecting V polarized ones.}
\end{figure}
This operation conserves the photon number in both input modes
without requiring measurements on the output of the two modes.
By applying this operation e.g. to the $H$ polarized components 
of two photonic qubits, it is then possible to realize a filter
that removes the $\mid\! H;V \rangle$ and $\mid\! V;H \rangle$
components while preserving coherence between $\mid\! H;H \rangle$
and $\mid\! V;V \rangle$. The complete setup is shown in 
figure \ref{filter}. By introducing an additional attenuation
in the vertically polarized path and by reversing the nonlinear
phase shift for the horizontally polarized components, the 
total filter effect of this setup becomes 
\begin{equation}
\label{eq:filter}
\hat{S}_{\mbox{filter}} = \frac{1}{4} \left(
\mid\! H;H\rangle\langle H;H\!\mid +  
\mid\! V;V\rangle\langle V;V\!\mid \right).
\end{equation}
The successful operation of this filter corresponds to 
a quantum nondemolition measurement of the relative alignment 
of $HV$ polarization. This measurement is not sensitive to 
any local polarization properties. The filter can therefore 
produce a variety of entangled output states \cite{Hof02}. 
Perhaps the most striking illustration of these nonlocal 
filter properties is the action
on a product state of a right circular polarized photon and
a left circular polarized photon,
\begin{eqnarray}
\label{eq:entangle}
\lefteqn{\hat{S}_{\mbox{filter}}\mid\! R;L\rangle =}
\nonumber \\ &&
\hat{S}_{\mbox{filter}} \;
\frac{1}{\sqrt{2}} 
\left(\mid\! H \rangle + i \mid\! V \rangle \right)
\otimes \frac{1}{\sqrt{2}} 
\left(\mid\! H \rangle - i \mid\! V \rangle \right)
\nonumber \\ &=& 
\frac{1}{8}(\mid\! H;H \rangle + \mid\! V;V \rangle)
\hspace{0.2cm}
= \frac{1}{8}(\mid\! R;L \rangle + \mid\! L;R \rangle).
\end{eqnarray}
The filter can thus be used to entangle photons originating
from separate and independent sources. 

While the quantum gates proposed in \cite{Kni01,Ral02}
and the quantum filter presented here and in \cite{Hof02}
allow a reliable implementation of all possible quantum 
operations, their main disadvantage is that they require 
not only a large number of beam splitters, but also 
an additional photon input of one extra photon per photonic
qubit, as well as a reliable detector to monitor the 
post-selection requirement. This effort is necessary so that the 
post-selection condition can be imposed without requiring 
any measurements in the output. However, more simple 
realizations of nonlinear interactions between photonic 
qubits are possible if the post selection ensuring the 
presence of exactly one photon in each qubit is performed
in the output instead. 

\section{Phase gate without additional input photons}

\begin{figure}
\footnotesize
\setlength{\unitlength}{0.58pt}
\begin{picture}(400,320)
%\put(0,0){\framebox(400,320){}}

%====V1
%--photon1in
\put(108,248){\line(-1,1){30}}
\put(112,252){\line(-1,1){30}}
\put(0,280){\makebox(80,20){Photon 1 in}}
%--polarizer1in
\put(120,220){\line(0,1){40}}
\put(120,260){\line(1,-1){20}}
\put(120,260){\line(-1,-1){20}}
\put(120,220){\line(1,1){20}}
\put(120,220){\line(-1,1){20}}
\put(105,230){\makebox(15,20){V}}
\put(120,230){\makebox(15,20){H}}
%--V1 path
\put(110,230){\line(-1,-1){90}}
\put(20,140){\line(0,1){7}}
\put(20,140){\line(1,0){7}}
\put(0,120){\makebox(30,20){loss}}
\put(110,90){\line(-1,1){70}}
%--left beam splitter
\put(20,185){\makebox(40,20){$R=1/3$}}
\put(40,135){\line(0,1){50}}

%--polarizer1out
\put(120,60){\line(0,1){40}}
\put(120,100){\line(1,-1){20}}
\put(120,100){\line(-1,-1){20}}
\put(120,60){\line(1,1){20}}
\put(120,60){\line(-1,1){20}}
\put(105,70){\makebox(15,20){V}}
\put(120,70){\makebox(15,20){H}}
%--photon1out
\put(108,72){\line(-1,-1){28}}
\put(112,68){\line(-1,-1){28}}
\put(80,40){\line(1,0){10}}
\put(80,40){\line(0,1){10}}
\put(0,20){\makebox(80,20){Photon 1 out}}

%====V2

%--photon2in
\put(292,248){\line(1,1){30}}
\put(288,252){\line(1,1){30}}
\put(320,280){\makebox(80,20){Photon 2 in}}
%--polarizer2in
\put(280,220){\line(0,1){40}}
\put(280,260){\line(-1,-1){20}}
\put(280,260){\line(1,-1){20}}
\put(280,220){\line(-1,1){20}}
\put(280,220){\line(1,1){20}}
\put(265,230){\makebox(15,20){H}}
\put(280,230){\makebox(15,20){V}}
%--V2 path
\put(290,230){\line(1,-1){90}}
\put(380,140){\line(0,1){7}}
\put(380,140){\line(-1,0){7}}
\put(370,120){\makebox(30,15){loss}}
\put(290,90){\line(1,1){70}}
%--right beam splitter
\put(340,185){\makebox(40,20){$R=1/3$}}
\put(360,135){\line(0,1){50}}

%--polarizer2out
\put(280,60){\line(0,1){40}}
\put(280,100){\line(-1,-1){20}}
\put(280,100){\line(1,-1){20}}
\put(280,60){\line(-1,1){20}}
\put(280,60){\line(1,1){20}}
\put(265,70){\makebox(15,20){H}}
\put(280,70){\makebox(15,20){V}}
%--photon2out
\put(292,72){\line(1,-1){28}}
\put(288,68){\line(1,-1){28}}
\put(320,40){\line(-1,0){10}}
\put(320,40){\line(0,1){10}}
\put(320,20){\makebox(80,20){Photon 2 out}}

%====H12
\put(270,90){\line(-1,1){140}}
\put(130,90){\line(1,1){140}}
%--main beam splitter
\put(180,185){\makebox(40,20){$R=1/3$}}
\put(200,135){\line(0,1){50}}

\end{picture}
\normalsize
\setlength{\unitlength}{1pt}
\vspace{-0.3cm}
\caption{\label{qgate} Schematic setup of a quantum phase gate
based on post-selection in the output. The nonlinear interaction
between the input photons is realized by the central beam 
splitter. No additional photon sources are required.}
\end{figure}
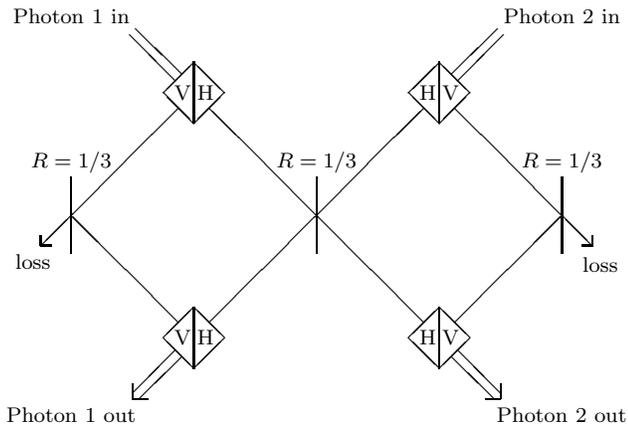

If post-selection in the output is allowed, it is 
possible to use both sides of the beam splitter to
realize an interaction between photons in modes 1 
and 2. If each mode has zero or one photon, the matrix
elements for the post-selected interaction are given 
by 
\begin{eqnarray}
\langle 0;0 \! \mid \hat{U}_R \mid \! 0;0 \rangle
&=& 1
\nonumber \\ 
\langle 0;1 \! \mid \hat{U}_R \mid \! 0;1 \rangle
&=& \sqrt{R}
\nonumber \\ 
\langle 1;0 \! \mid \hat{U}_R \mid \! 1;0 \rangle
&=& \sqrt{R}
\nonumber \\ 
\langle 1;1 \! \mid \hat{U}_R \mid \! 1;1 \rangle
&=& 2R-1.
\end{eqnarray}
This diagonal four by four matrix already describes the
basic function of a phase gate since the two photon
term is an interference between mutual reflection and
mutual transmission of the indistinguishable photons.
By choosing a reflectivity of $R=1/3$, the non-unitary 
amplitude factors correspond to a linear attenuation,
\begin{eqnarray}
\label{eq:1/3}
\langle 0;0 \! \mid \hat{U}_{1/3} \mid \! 0;0 \rangle
&=& \hspace{0.5cm} 1
\nonumber \\ 
\langle 0;1 \! \mid \hat{U}_{1/3} \mid \! 0;1 \rangle
&=& \sqrt{1/3}
\nonumber \\ 
\langle 1;0 \! \mid \hat{U}_{1/3} \mid \! 1;0 \rangle
&=& \sqrt{1/3}
\nonumber \\ 
\langle 1;1 \! \mid \hat{U}_{1/3} \mid \! 1;1 \rangle
&=& - 1/3.
\end{eqnarray}
The total setup for a pair of photonic qubits is then 
obtained by applying (\ref{eq:1/3}) to one of the 
polarization components of each input qubit\cite{Hof01,Ral01}.
This very compact and symmetric setup is illustrated in
figure \ref{qgate}.
Its function is described by the operator 
$\hat{S}_{\mbox{qpg}}$ with
\begin{eqnarray}
\hat{S}_{\mbox{qpg}} \mid \! V;V\rangle &=& 
\hspace{0.5cm} \frac{1}{3} \mid \! V;V\rangle
\nonumber \\
\hat{S}_{\mbox{qpg}} \mid \! V;H\rangle &=& 
\hspace{0.5cm} \frac{1}{3} \mid \! V;H\rangle
\nonumber \\
\hat{S}_{\mbox{qpg}} \mid \! H;V\rangle &=& 
\hspace{0.5cm} \frac{1}{3} \mid \! H;V\rangle
\nonumber \\
\hat{S}_{\mbox{qpg}} \mid \! H;H\rangle &=& 
\hspace{0.2cm} -  \frac{1}{3} \mid \! H;H\rangle.
\end{eqnarray}
Since this quantum gate requires no additional
photon sources or photon detectors, it appears to be
a most promising candidate for the realization of
networks for multi-qubit operations. 

\section{Conclusions}
As the examples given above clearly demonstrate, 
post-selection methods can be very useful in 
realizing quantum operations that are difficult to 
obtained from a direct physical interaction in a dissipation
free environment. Instead of controlling the Hamiltonian,
post-selection applies the dynamics of quantum measurement
to realize the desired coherent interactions. 
It is then possible to employ the same technologies 
originally developed for an improvement of measurement
precision to the evolution of the quantum state.
In optics, this method can be applied to solve the problem
of realizing strong nonlinear interactions between individual
photons. As shown above, the nonlinearity necessary for
such interactions is already present in the photon number
conserving subspace of a conventional beam splitter. This
nonlinearity can be post-selected by reliably controlling 
the photon numbers in both the input and the output. 
Effectively, the nonlinearity is implemented by precise photon 
counting measurements. Using recently developed photon detection
technologies \cite{Tak99,Kim99}, it should then be 
possible to perform arbitrary quantum operations on photonic 
qubits with only moderate rates of error.

%=========================================================

%=========================================================

\end{document}